\documentclass{article}
\usepackage{spconf,amsmath,epsfig}

\title{USING A PITCH-SYNCHRONOUS RESIDUAL CODEBOOK FOR HYBRID HMM/FRAME SELECTION SPEECH SYNTHESIS}
\twoauthors
  {Thomas Drugman, Alexis Moinet, Thierry Dutoit}
	{Facult\'e Polytechnique de Mons\\ TCTS Lab\\ 31, Boulevard Dolez\\7000 Mons, Belgium}
  {Geoffrey Wilfart}
	{Acapela Group\\
	Research and Development\\
	33, Boulevard Dolez\\7000 Mons, Belgium}
	
\begin{document}
%
\maketitle
\begin{abstract}
This paper proposes a method to improve the quality delivered by statistical parametric speech synthesizers. For this, we use a codebook of pitch-synchronous residual frames, so as to construct a more realistic source signal. First a limited codebook of typical excitations is built from some training database. During the synthesis part, HMMs are used to generate filter and source coefficients. The latter coefficients contain both the pitch and a compact representation of target residual frames. The source signal is obtained by concatenating excitation frames picked up from the codebook, based on a selection criterion and taking target residual coefficients as input. Subjective results show a relevant improvement compared to the basic technique.   
\end{abstract}

\section{Introduction}\label{sec:intro}
Two text-to-speech technologies have clearly emerged these last years. On one hand, the Unit Selection method \cite{Hunt} concatenates speech units picked up from a very large corpus, avoiding signal processing manipulations as much as possible, in order to minimize segmental quality degradations. Its biggest drawbacks lie in the difficulty of producing voice quality variations, required to produce expressive speech, and in the limited voice modification/conversion that are allowed.

On the other hand, Statistical Parametric Speech Synthesis \cite{Black} models the speech signal in various contextual situations. HMM-based synthesizers \cite{Tokuda} have thus recently gained considerable attention for their flexibility, smoothness and small footprint. Nevertheless their main disadvantage is the quality of the produced speech, which exhibits the typical \emph{buzziness} found in the old LPC-based speech coders. While techniques for modeling the filter are rather well-established, it is not the case for the source representation. In order to overcome this hindrance, some works have proposed a more complex excitation model. In 2001 Yoshimura et al. \cite{Yoshimura} presented the use of a Mixed Excitation (ME) in HMM-based speech synthesis. The ME is expressed as a weighted sum of both filtered pulse sequence and white noise. Gonzalvo et al. also recently confirmed the  efficiency of this technique for Spanish synthesis \cite{Gonzalvo}. In a similar way, Maia et al. \cite{Maia} proposed a ME consisting of a set of state-dependent filters derived through a closed-loop procedure.

This paper also aims at reducing the produced buzziness by adopting a more subtle source signal. Contrarily to previous approaches we believe that the best way to achieve this is to use segments of \textbf{real} excitation. We propose a method using a small codebook of pitch-synchronous residual frames. Assuming that the residual obtained by inverse filtering approximates the glottal flow first derivative, it is reasonable to consider that a limited codebook of typical excitations can be built. Futhermore applying time scaling operations on these frames should preserve the most important features of the glottal pulse (such as the open quotient and asymmetry coefficient) \cite{Cabral1}.

During the synthesis part, HMMs are used to generate filter and source coefficients. The latter coefficients contain both $F_0$ and a compact representation of target residual frames. The source signal is then constructed by overlap-adding pitch-sized residual frames (picked up from the codebook on basis of the residual coefficients) during voiced regions. The synthesis process is therefore comparable to a PS-CELP technique \cite{PS-CELP}. Our system can thus be viewed as an hybrid synthesizer relying on a statistical modeling of both the source and filter coefficients, and using a residual frame selection for producing the source excitation from the target source parameters generated by the statistical model.

Section \ref{sec:codebook} shows how we build and compress our residual codebook. A criterion for residual selection, based on a low-frequency representation, as well as the source signal generation technique are presented in Section \ref{sec:residual}. In Section \ref{sec:AS} we validate our method on an Analysis-Synthesis task. Section \ref{sec:synth} implements the approach in a Text-to-Speech framework and gives results. Section \ref{sec:conclu} concludes and presents some guidelines for future works.

\section{Codebook construction}
\label{sec:codebook}

Our method requires to build a codebook of pitch-synchronous residual frames. Spectral analysis is first performed so as to obtain residuals, using Mel-Generalized Cepstral coefficients (MGC). Our system performs MGC analysis with $\alpha$=0.42 ($F_s$=16kHz) and $\gamma$=-1/3, as suggested in \cite{Blizzard06}. Residual signals are then obtained by inverse filtering.

\begin{figure}[!ht]
  \centering
  \includegraphics[width=0.4\textwidth]{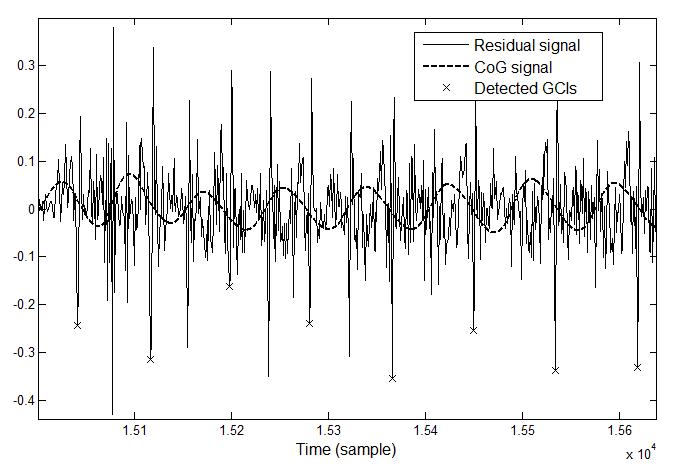}
  \caption{GCI detection disambiguation thanks to the CoG signal}
  \label{fig:GCI}
\end{figure}

Locating the Glottal Closure Instants (GCI) then allows to isolate GCI-centered two-period long excitation frames. GCIs are particular events in the speech signal corresponding to the sudden return of the vocal folds. Most of the time they can be easily detected by applying peak picking on the residual signal. Nonetheless some cases (such as nasals) are more complex. Inspired from \cite{Kahawara}, we compute the Center of Gravity (CoG) in energy on two-period long speech frames. Zero-crossings (from positve to negative) of this signal, associated with a peak picking technique, remove possible ambiguities on the GCI location determination (Figure \ref{fig:GCI}). The codebook is then filled with GCI-centered, two-period long and Hanning-windowed residual frames. 

\begin{figure}[!ht]
  \centering
  \includegraphics[width=0.45\textwidth]{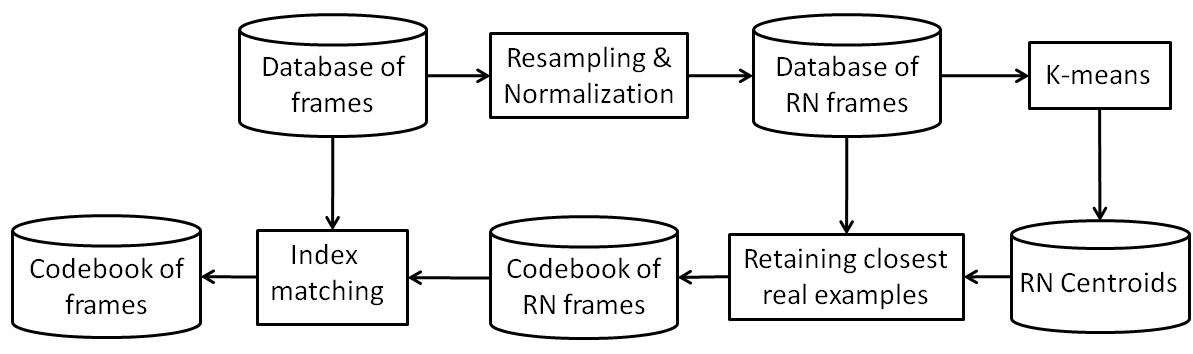}
  \caption{Codebook compression workflow}
  \label{fig:Compression}
  \vspace{-12pt} 
\end{figure}

Preserving the small footprint of the HMM-based synthesizer is crucial for industrial applications. It is consequently useful to compress our codebook (Figure \ref{fig:Compression}). The first step aims at representing the residuals in a compact way so as to make them amenable to clustering. For this, we used a Resampled and Normalized (RN) version of the residual frames. This step is tied with the selection criterion and is further explained in Section \ref{ssec:selection}. In our system we applied the K-Means algorithm on these RN frames, retaining typically 100 centroids (we made trials with various codebook sizes and found that 100 centroids were enough for keeping the compression almost inaudible, see Section \ref{sec:AS}). Indeed as the variability due to formants and pitch has been eliminated a great gain of compression can be expected. A real residual frame has then to be assigned to each centroid. For this we keep in mind the difficulties that will appear when the residual frame will have to be converted back to targeted pitch frames (see Section \ref{ssec:modifications}). In order to reduce the appearance of ``energy holes'' during the synthesis, frames composing the compressed inventory are chosen so as to exhibit a pitch as low as possible. For each centroid we therefore take the N-closest frames (according to their RN (i.e euclidian) distance, N=10 here) and only retain the longest frame.

\section{Source signal generation}
\label{sec:residual}

\begin{figure}[!ht]
  \centering
  \includegraphics[width=0.45\textwidth]{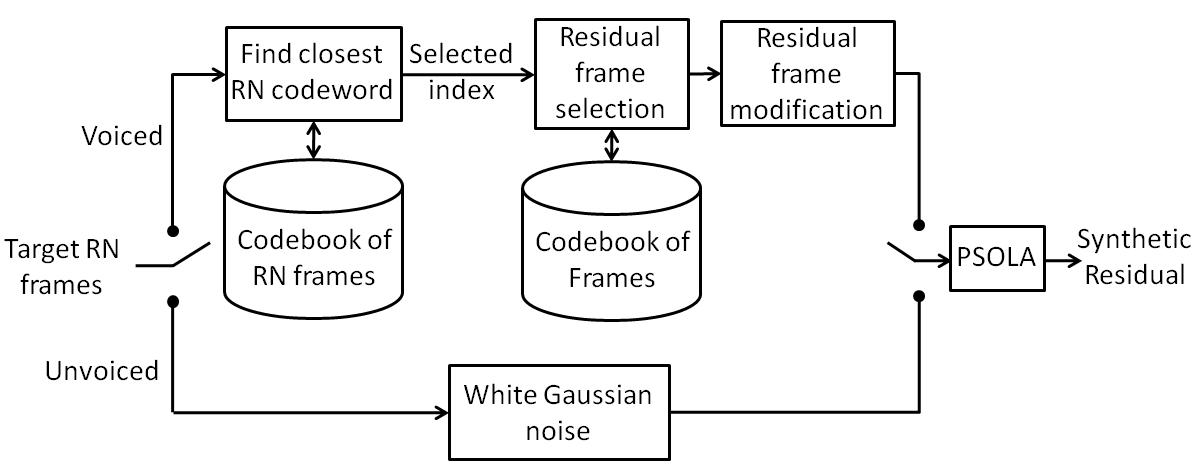}
  \caption{The source signal construction from the target RN source parameters.}
  \label{fig:Synthesis}
  \vspace{-12pt}
\end{figure}

Once the codebooks are available, the source signal used in the sythesis part is achieved as in Figure \ref{fig:Synthesis}. Unvoiced excitation is modeled by white noise. Given target RN source frames (Section \ref{ssec:selection}), voiced residual frames are selected from the codebook and then modified to exhibit the target prosody (see Sections \ref{ssec:selection} and \ref{ssec:modifications}). The source signal is then obtained by applying PSOLA to the residual frames.

\subsection{Residual frame selection}\label{ssec:selection}
\vspace{-4pt}
How to define a similarity measure between two residual frames? Although this problem has been extensively studied for speech, it has to be re-analyzed for the residual signal. We here consider that the main characteristics of the signal lie in the overall shape of the residual waveform. For this, frames are Resampled on 20 coefficients and Normalized in energy (hereafter called RN frames). These 20 samples correspond to a low-frequency signature of the underlying residual frames. In earlier experiments we also attempted to use both FFT amplitudes and phases, as well as group delays. However the RN technique gave the most convincing results.
By computing a Mean Square Error on the RN frames, it is possible to select the closest residual frames (with their original prosodic features) from the codebook.

\subsection{Residual frame modification}\label{ssec:modifications}
Once a frame has been selected from the codebook, its prosody still needs to be adapted to target prosody. Residual frames selected from the codebook are thus converted to the target pitch (by resampling) and energy. Indeed, as previously mentioned, since the residual is assumed to approximate the glottal flow first derivative, resampling the residual frames by interpolation and decimation preserves their shape and consequently their most important features (notably the open quotient and asymmetry coefficient). To avoid the emergence of energy holes in high frequencies, care is taken when compressing the codebook (see Section \ref{sec:codebook}), although some solutions have been proposed in \cite{Cabral2}.

\vspace{-2pt}
\section{Analysis-Synthesis experiments}
\label{sec:AS}
\vspace{-4pt}

A preliminary step was required to verify the effectiveness of our method. We have first applied the above mentioned codebook construction on a training dataset, as in Section \ref{sec:codebook}. The test sentences (not contained in the dataset) were then analyzed. GCIs were detected such that the framing is GCI-centered and two-period long during voiced regions. To make the selection possible, these frames were resampled and normalized so as to get the RN frames. These latter frames were input into the source signal reconstruction workflow shown in Figure \ref{fig:Synthesis}.

Once selected from the codebook, each residual frame was modified in pitch and energy so as to replace the original one. Unvoiced segments were replaced by a white noise segment of same energy. The resulting source signal was then filtered by the original MGC coefficients previously extracted.

\begin{figure}[!ht]
  \centering
  \includegraphics[width=0.45\textwidth]{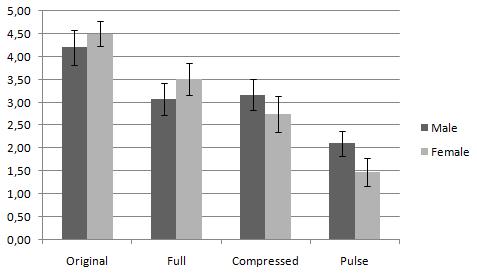}
  \caption{MOS test results for the Analysis-Synthesis Experiments (with their $95\%$ confidence intervals).}
  \label{fig:AS_MOS}
\end{figure}

Experiments were performed on 5 voices (2 females, 3 males) from the CMU ARCTIC database publicly available in \cite{ARCTIC} (speakers AWB, BDL, CLB, JMK and SLT). For each speaker four samples were assessed: the original file and the resynthesized speech using the full (whole database of residual frames) and compressed codebooks, or using a pulse sequence during voiced excitation (i.e the basic technique used in HMM-based synthesis). Note that even for the latter technique, GCI-synchronous pulses were used so as to capture micro-prosody (the resulting vocoded speech therefore provided a high-quality baseline). The results of the MOS test submitted to 20 listeners are shown in Figure \ref{fig:AS_MOS}. A relevant improvement compared to the pulse method can be noticed. Compressing the codebook for male speakers turns out not to have any effect while some loss of quality is observed for female speakers. These trends are noticeable among all speakers.

\vspace{-4pt}

\section{Synthesis framework}
\label{sec:synth}
\vspace{-4pt}
\subsection{Framework}\label{ssec:Sframework}

\begin{figure}[!ht]
  \centering
  \includegraphics[width=0.4\textwidth]{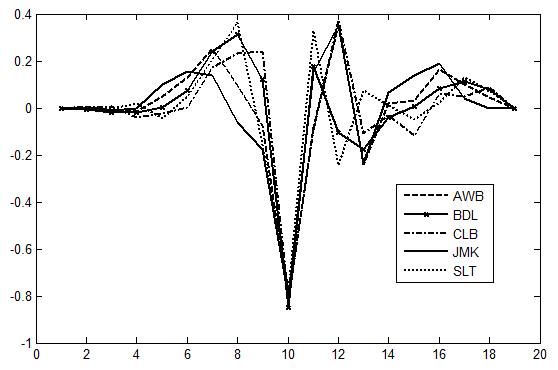}
  \caption{The first eigen-RNframe for five speakers from the CMU ARCTIC database.}
  \label{fig:PCA}
\end{figure}

We also tested our method on a HMM-based speech synthesizer. As explained in Section \ref{ssec:selection}, the selection criterion that we use here is related to the choice of the RN frames for compactly describing residual frames. Nevertheless these coefficients are not suited for a statistical modeling/generation for the following reasons:
\begin{itemize}
	\item they are highly correlated (whereas common modelings make use of diagonal covariance matrices),
	\item they do not have good interpolation properties.
\end{itemize}
To overcome these drawbacks, the RN coefficients are linearly transformed by a Principal Component Analysis (PCA) before being modeled by HMMs. No dimensionality reduction is applied here. Figure \ref{fig:PCA} displays the first eigen-RNframe for the five speakers analyzed in Section \ref{sec:AS}. A strong similarity with Liljencrants/Fant-like models \cite{LF} of the glottal source can be observed. 

Simultaneous modeling of all parameters is performed using the HMM-based Speech Synthesis System (HTS, \cite{Tokuda}). The filter parameters consist of the MGC-LSP coefficients while the source parameters comprise both pitch and PCA coefficients. Since these coefficients are only meaningful in voiced regions, we opted for a Multi-Space Distribution (MSD) \cite{MSD}. PCA parameters are therefore seen as an additional 3 MSD-streams, made of the static vectors, their first and second derivatives, for a total of 7 coefficient streams.

At synthesis time, HMMs generate sequences of filter and source parameters. PCA coefficients are converted back to RN frames so as to be the input of the source signal construction stage (see Figure \ref{fig:Synthesis}). Synthesized speech is finally obtained by feeding the Mel-Generalized Log Spectral Approximation (MGLSA) filter with our excitation signal.

\subsection{Results}\label{ssec:Sresults}
\begin{figure}[!ht]
  \centering
  \includegraphics[width=0.45\textwidth]{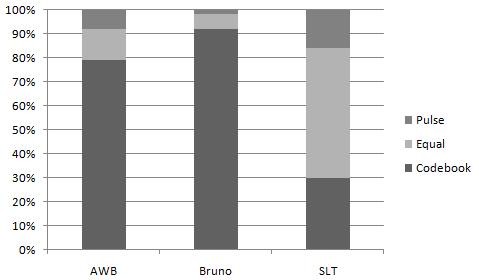}
  \caption{Preference test for three speakers.}
  \label{fig:Pref}
  \vspace{-12pt}
\end{figure}

Synthesis was performed for three speakers: AWB (Scottish male), Bruno (French male, kindly provided by the Acapela Group) and SLT (US female). The duration of the training set was about 50 min for AWB and SLT, and 2 h for Bruno. A preference test (3 sentences per speaker) between our method and the basic approach using pulse sequences was submitted to 20 subjects (Figure \ref{fig:Pref}). A significative gain is noticed for male speakers while results are more mitigated on SLT. Solving the latter observation is the object of ongoing research.

\section{Conclusions and future work}
\label{sec:conclu}

This paper proposed a technique for improving the quality delivered by statistical parametric speech synthesizers, by making use of a codebook of pitch-synchronous residual frames. A significant improvement with regard to the basic technique has been observed through MOS tests, while the system footprint remains under one Mb. As future work, we plan to use residual units with larger durations, for instance by employing sequences of consecutive frames instead of single frames. As our current framework only takes local decisions, the definition of a concatenation cost besides the selection one should also be investigated. 

\section{Acknowledgments}\label{sec:Acknowledgments}

Thomas Drugman is supported by the ``Fonds National de la Recherche
Scientifique'' (FNRS).

\bibliographystyle{IEEEbib}
\bibliography{strings,refs}

\end{document}